\documentclass[aip,jmp,amssymb, 12pt]{revtex4-1}
\usepackage{graphicx}
\usepackage{amsmath}
\usepackage{geometry}
\usepackage{txfonts}
\usepackage{comment}
\usepackage{amssymb}
\usepackage{makeidx}
\usepackage{cancel}
\usepackage[usenames,dvipsnames,svgnames]{xcolor}

\def\Tr{\mathrm{Tr}}

\def\dd{\mathrm{d}}
\def\im{\mathrm{Im}}

\newcommand \be{\begin{eqnarray}}
\newcommand \ee{\end{eqnarray}}
\newcommand{\del}{\partial}

\begin{document}

\title{Diffusion in the space of complex Hermitian matrices - microscopic properties of the averaged characteristic polynomial and the averaged inverse characteristic polynomial}

\begin{abstract} 
We show that the averaged characteristic polynomial and the averaged inverse characteristic polynomial, associated with Hermitian matrices whose elements perform a random walk in the space of complex numbers, satisfy certain partial differential, diffusion-like, equations. These equations are valid for matrices of arbitrary size. Their solutions can be given an integral representation that allows for a simple study of their asymptotic behaviors for a broad range of initial conditions.  
\end{abstract}

\author{Jean-Paul Blaizot}
\email{Jean-Paul.Blaizot@cea.fr} \affiliation{IPhT, CEA-Saclay,
91191 Gif-sur Yvette, France}

\author{Jacek Grela} \email{grela@th.if.uj.edu.pl} \affiliation{M.
Smoluchowski Institute of Physics,  Jagiellonian University,
PL--30--059 Cracow, Poland}

\author{Maciej A. Nowak}
\email{nowak@th.if.uj.edu.pl} \affiliation{M. Smoluchowski Institute
of Physics and Mark Kac Center for Complex Systems Research,
Jagiellonian University, PL--30--059 Cracow, Poland}

\author{Piotr Warcho\l{}} \email{piotr.warchol@uj.edu.pl} \affiliation{M.
Smoluchowski Institute of Physics,  Jagiellonian University,
PL--30--059 Cracow, Poland}
\date{\today}


\maketitle

\section{Introduction}
As two of us have argued\cite{BN2}, a particularly interesting, hydrodynamic-like picture of the evolution of the averaged characteristic polynomial (hereafter ACP) emerges if one exploits Dyson's idea \cite{D1} of introducing temporal dynamics into random matrix ensembles. 
Consider for instance a Hermitian, $N\times N$ matrix whose entries perform a properly normalized, continuous
random walk in the space of complex numbers. Let this process be initiated with a matrix filled with zeros.  For this particular choice it was shown that the ACP is
equal to a certain time dependent monic Hermite polynomial in a complex variable $z$,  that satisfies a complex diffusion equation with a diffusion constant equal to $-\frac{1}{2N}$. Moreover, the ACP can be transformed into a different function (by taking its logarithmic derivative with respect to the complex variable $z$) that is governed by the viscid Burgers equation. In the large $N$ (inviscid) limit, the latter admits solutions exhibiting shocks whose positions coincide with the edges of the eigenvalue spectrum. In the finite $N$ (viscid) case one can perform an expansion around the shocks to obtain the well known Airy asymptotic behavior of the averaged characteristic polynomial. Note finally, that this analysis can be carried over to the case of diffusing Wishart matrices\cite{BNW1}. 

In this paper we demonstrate the robustness of the above ideas. To this end we show that both the ACP  and the averaged inverse characteristic polynomial (AICP) satisfy the same complex diffusion equation, except that for the former the diffusion constant is equal to $-\frac{1}{2N}$, whereas for the latter it is $\frac{1}{2N}$. The new proof works regardless of the actual form of the initial condition imposed on the process and this allows us to examine two different, generic scenarios. The solution of the diffusion equation leads to simple integral representations for both the ACP and the AICP, which makes it possible to  study their asymptotic, large $N$, universal behaviors. In particular, we recover the known scaling property of the ACP and the AICP at the edge of the spectrum in terms of Airy functions. Moreover, for a process initiated with at least two distinct eigenvalues, when two edges of the spectrum meet, the Pearcey functions emerge. Thus, the present diffusion scenario provides a natural and simple way to re-derive the universal functions corresponding to the fold and cusp singularities in random matrix models.

The paper is organized as follows. We start by introducing the stochastic evolution of the studied matrix. Using the representations of the determinant and its inverse as Gaussian integrals over, respectively, Grassman or complex variables, we derive the diffusion equations for the ACP and the AICP. For the simplest scenario, in which the process is initialized with a null matrix, we crosscheck the equation for the AICP by exhibiting the equivalence of its solution with the Cauchy transform of the ACP. In the following section, we derive the corresponding Burgers equation, which we solve in the large $N$ limit with the method of complex characteristics and obtain the associated Green's function for two different generic examples. In the first one, the initial matrix is filed with zeros, whereas in the second, it has two distinct non vanishing eigenvalues. We subsequently use the saddle point method to inspect how the ACP and AICP behave in the former scenario, at the points corresponding to the edges of the probability density function for the eigenvalues, asymptotically when $N$ is increased. Next, we turn to the case of the latter initial condition. The spectrum then forms two disjoint lumps of eigenvalues that eventually collide. We perform the saddle point analysis around the time and point of this collision. Finally, we mention how the ACP and AICP can be used to reconstruct the random matrix kernel for the problem, the details are however left for the appendix.
The last section summarizes our conclusions.

\section{Diffusion of Hermitian matrices}
Let us introduce an $N\times N$ Hermitian matrix $H$ by defining its complex entries according to:
\begin{align}
H_{ij}= \left\{ 
	\begin{array}{l r}
	x_{ii}, & i=j, \\
	\frac{1}{\sqrt{2}} ( x_{ij}+i y_{ij} ), & i\neq j,
	\end{array}\right.
\end{align}
where $x_{ij}=x_{ji}$ and $y_{ij}=-y_{ji}$, with $x_{ij}$ and $y_{ij}$ real. Furthermore let $x_{ij}$ and $y_{ij}$ perform white noise driven, independent random walks, such that
\begin{align}
	\left \langle\delta H_{ij} \right\rangle = 0, \qquad \left \langle \left(\delta H_{ij}\right)^{2}\right\rangle=\frac{1}{N}\delta \tau
\end{align} 
for any $i$ and $j$. Let $P(x_{ij},\tau)P(y_{ij},\tau)$ be the probability that the off diagonal matrix entry $H_{ij}$  will change from its initial state to $\frac{1}{\sqrt{2}} ( x_{ij}+i y_{ij} )$ after time $\tau$. Analogically, $P(x_{ii},\tau)$ is the probability of the diagonal entry $H_{ii}$ becoming equal to $x_{ii}$ at $\tau$.
The evolution of these functions is governed by the following diffusion equations:
\begin{align}
\label{entrydiff}
\frac{\del}{\del \tau}P(x_{ij},\tau) & = \frac{1}{2N}\frac{\del^{2}}{\del x_{ij}^{2}}P(x_{ij},\tau) , \nonumber \\
\frac{\del}{\del \tau}P(y_{ij},\tau) & = \frac{1}{2N}\frac{\del^{2}}{\del y_{ij}^{2}}P(y_{ij},\tau) , \quad i\ne j.
\end{align}
Moreover, the joint probability density function
\begin{align}
	 \qquad P(x,y,\tau)\equiv\prod_{k}P(x_{kk},\tau)\prod_{i<j} P(x_{ij},\tau)P(y_{ij},\tau)\label{prob}
\end{align}
satisfies  the following Smoluchowski-Fokker-Planck equation
\begin{align}
\del_\tau P(x,y,\tau)=\mathcal{A}(x,y) P(x,y,\tau), \qquad \mathcal{A}(x,y) = \frac{1}{2N}\sum_k \frac{\del^{2}}{\del x_{kk}^{2}}+\frac{1}{2N}\sum_{i<j} \left ( \frac{\del^{2}}{\del x_{ij}^{2}}+ \frac{\del^{2}}{\del
y_{ij}^{2}} \right ). \label{prob1}
\end{align}
With the setting thus defined, let us proceed to the derivation of the partial differential equations governing the ACP and AICP.

\subsection{Evolution of the averaged characteristic polynomial}
Let  $\pi_{N}(z,\tau)$ be the averaged characteristic polynomial associated with the diffusing matrix $H$: $\pi_{N}(z,\tau)\equiv\langle \det\left(z-H\right)\rangle$, where the angular brackets denote the averaging over the time dependent probability density (\ref{prob}). In order to derive the partial differential equation governing the ACP, we write the determinant as a Gaussian integral over  Grassmann variables $\eta_i, \bar \eta_i$:
\begin{align}
{\rm det}~{A} = {\int {{\prod _{i,{j}}{{\rm d}{\eta }_{{i}}}}{{\rm d}{{{\overline\eta } }_{{j}}}}}}\,{\exp{\left({\overline{{\eta }_{{i}}}{A}_{{ij}}{\eta
}_{{j}}}\right)}}.\label{gras}
\end{align}
This allows us  to express the averaged characteristic polynomial in the following way:
\begin{align}
\pi_{N}(z,t) & = \int \mathcal{D}[\bar{\eta},\eta,x,y]P(x,y,\tau)\exp{\left[\bar{\eta}_i\left(z\delta_{ij}-H_{ij}\right)\eta_{j}\right]},\label{repr}
\end{align}
where the joint integration measure is defined by
\begin{align}
\mathcal{D}[\bar{\eta},\eta,x,y] & \equiv \prod _{i,j}{\rm d}\eta_{i}{\rm d}\overline{\eta}_{j} \prod_{k}{\rm d}x_{kk}\prod_{n<m}{\rm d}x_{nm}{\rm
d}{y}_{nm}.
\end{align}
The Hermicity condition ($H_{ij}=\bar{H}_{ji}$) allows us to write the argument of the exponent of (\ref{repr}) in a convenient form:
\begin{align}
T_g(\bar{\eta},\eta,x,y,z) & \equiv\sum_{r}\bar{\eta}_{r}\left(z-x_{rr}\right)\eta_{r}
-\frac{1}{\sqrt{2}}\sum_{n<m}\left[
x_{nm}\left(\bar{\eta}_{n}\eta_{m}-\eta_{n}\bar{\eta}_{m}\right)+iy_{nm}\left(\bar{\eta}_{n}\eta_{m}+\eta_{n}\bar{\eta}_{m}\right)\right].\nonumber
\end{align}
Note that the time dependence of $\pi(z,\tau)$ resides entirely in $P(x,y,\tau)$. By differentiating Eq.~(\ref{repr}) with respect to $\tau$, and using Eq.~(\ref{prob1}), one ends up with an expression where the operator $\mathcal{A}(x,y)$ acts on the joint probability density function. Integrating by parts with respect to $x_{ij}$ and $y_{ij}$, one obtains
\begin{align}
\del_{\tau}\pi_{N}(z,\tau) & = \int \mathcal{D}[\bar{\eta},\eta,x,y]P(x,y,\tau) \mathcal{A}(x,y) \exp{\left[T_g(\bar{\eta},\eta,x,y,z)\right]}.
\end{align} 
At this point, we differentiate with respect to the matrix elements (acting with $\mathcal{A}(x,y)$), exploit  some simple properties of Grassmann variables, and obtain:
\begin{align}
\del_{\tau}\pi_{N}(z,\tau) = -\frac{1}{N} \int \mathcal{D}[\bar{\eta},\eta,x,y]P(x,y,\tau)
\sum_{i<j}\bar{\eta}_i \eta_i \bar{\eta}_j \eta_{j} \exp{\left[T_g(\bar{\eta},\eta,x,y,z)\right]}.
\end{align}
It is easily verified that this expression, when multiplied by $-2N$, matches the double differentiation with respect to $z$ of Eq.~(\ref{repr}).
We thus end up with
\begin{align}
\label{eq:acp}
\del_{\tau}\pi_{N}(z,t)=-\frac{1}{2N}\del_{zz}\pi_N(z,\tau).
\end{align}
This is the sought for diffusion equation for the ACP. Note that the same equation was already obtained in Ref.~\cite{BN2}, albeit for a very specific initial condition, for which $\pi_{N}(z,t)$ is a scaled Hermit polynomial. The present derivation has the advantage of being independent of the choice of initial condition. 
\subsection{Evolution of the averaged inverse characteristic polynomial}
We now turn to the averaged inverse characteristic polynomial 
\begin{align}
\theta_{N}(z,t)\equiv\left\langle \frac{1}{\det\left(z-H\right)}\right\rangle, \label{aicp0}
\end{align}
to which we are going to apply a similar strategy. 
In this case, we use he fact that the inverse of a determinant has a well-known representation in terms of a Gaussian integral over complex variables $\xi_{i}$:
\begin{align}
	\frac{1}{{\rm det}{A}}=\int \prod _{i,j}{\rm d}\xi_{i}{\rm d}{\overline\xi}_{j}\exp{\left(-\overline{\xi}_{i}A_{ij}\xi_{j}\right)}.\label{compl}
\end{align}
As in the ACP case, we use this representation to express (\ref{aicp0}) as
\begin{align}
\label{aicpsol}
\theta_{N}(z,\tau) = \int \mathcal{D}[\bar{\xi},\xi,x,y]P(x,y,\tau)\exp{\left[\bar{\xi}_i\left(H_{ij}-z\delta_{ij}\right)\xi_{j}\right]},
\end{align}
where, again, the proper notation for the joint integration measure was introduced. Performing the differentiation with respect to $\tau$ yields:
\begin{align}
\del_{\tau}\theta_{N}(z,\tau) & = \int \mathcal{D}[\bar{\xi},\xi,x,y]P(x,y,\tau) \mathcal{A}(x,y) \exp{\left[T_c(\bar{\xi},\xi,x,y)\right]}
,\end{align}
with
\begin{align}
T_c(\bar{\xi},\xi,x,y,z) & \equiv \sum_{r}\bar{\xi}_{r}\left(x_{rr}-z\right)\xi_{r}
+\frac{1}{\sqrt{2}}\sum_{n<m}\left [
x_{nm}\left(\bar{\xi}_{n}\xi_{m}+\xi_{n}\bar{\xi}_{m}\right)+iy_{nm}\left(\bar{\xi}_{n}\xi_{m}-\xi_{n}\bar{\xi}_{m}\right)\right ], \nonumber
\end{align}
where we have used (\ref{prob1}), the hermicity of $H$ and we have performed integrations by parts. After differentiation with respect to the matrix elements, one obtains:
\begin{align}
\del_{\tau}\theta_{N}(z,\tau) = \frac{1}{N} \int \mathcal{D}[\bar{\xi},\xi,x,y]P(x,y,\tau)\left(
\sum_{i<j}\bar{\xi}_i \xi_i \bar{\xi}_j \xi_{j}+\frac{1}{2}\sum_{k}\bar{\xi}_k\xi_k\bar{\xi}_{k}\xi_k \right) \exp{\left[ T_c(\bar{\xi},\xi,x,y,z)\right]}, 
\end{align}
which, multiplied by $2N$, matches the double differentiation of Eq.~(\ref{aicpsol})  with respect to $z$.
The  final result reads
\begin{align}
\del_{\tau}\theta_{N}(z,t)=\frac{1}{2N}\del_{zz}\theta_N(z,t),
\label{aicp}
\end{align}
the announced diffusion equation for the AICP.
\section{The integral representation}
The main advantage of the equations derived above is that they have obvious solutions in terms of initial condition dependent integrals. 
In this section we explicitly state those representations and show additionally how, for the simplest initial condition, one is a Cauchy transform of the other. 
Let us also note here that these types of integrals were obtained\cite{BK2} as representations of multiple orthogonal polynomials\cite{DF1,BK1} and equivalently
as averaged characteristic polynomials of GUE matrices perturbed by a source\cite{PFOR}.

\subsection{The averaged characteristic polynomial}


One can verify by a direct calculation that the expression 
\begin{align}
\label{acpsol}
\pi_N(z,\tau)= \mathcal{C} \,\tau^{-1/2} \int_{-\infty}^{\infty} \exp{\left(-N\frac{(q-iz)^{2}}{2\tau}\right)}\,\pi_N(-iq,\tau=0)\,\dd q,
\end{align}
satisfies the complex diffusion equation (\ref{eq:acp}) governing the evolution of the averaged characteristic polynomial. The imaginary unit in the exponent and in the argument of the initial condition arises from the negative value of the diffusion constant in this equation. For finite $N$, the most general form of the initial condition is $\pi_N(z,\tau=0) = \prod_i (z-\lambda_i)$, where the $\lambda_i$'s are real due to the Hermiticity of the initial matrix $H(\tau=0)$.
%
Exploiting  the steepest descent method to match this with Eq. (\ref{acpsol}), one determines the constant term $\mathcal{C}$. The saddle point associated with $\tau\to 0$ is $u_0 = iz$.
 Performing the Gaussian integration around the saddle point, we obtain $\mathcal{C}=\sqrt{\frac{N}{2\pi}}$ so that Eq.~(\ref{acpsol}) reads
\begin{align}\label{integralrepres3}
\pi_N(z,\tau)= \sqrt{\frac{N}{2\pi\tau}}\int_{-\infty}^\infty \exp{\left(-N\frac{(q-iz)^{2}}{2\tau}\right)}\,\pi_{N}(-iq,\tau=0)\,\dd q.
\end{align}

\subsection{The averaged inverse characteristic polynomial}
The integral representation of the averaged inverse characteristic polynomial arising as a solution to the partial differential equation (\ref{aicp}) is
\begin{align}
\theta_N(z,\tau)= \mathcal{C}\int_{\Gamma} ~ \exp{\left(-N\frac{(q-z)^{2}}{2\tau}\right)}\,\theta_{N}(q,\tau=0)\,\dd q . \label{aicpsol1}
\end{align}
As in the case of the ACP, the initial condition has to be recovered. Here however, $\theta_N(z,\tau=0)$ has poles on the real axis, and the contour $\Gamma$ must avoid these poles. A first possibility is to  choose $\Gamma_+$ parallel and slightly above  the real axis. In this case the saddle point analysis for $\tau\to 0$ is performed by moving $\Gamma_+$ upward so that it crosses the saddle point $q_0=z$. Obviously, this is possible only if ${\rm Im}z>0$. If instead ${\rm Im}z<0$, we need to choose $\Gamma_-$ also  parallel to the real axis but slightly below. Imposing an integration contour that would switch from the upper to the lower half plane (and {\it vice versa}) in between the poles, would results in a function no longer being the solution of the initial problem. 

In the simple case of $\pi_N(z,\tau=0) = z^N$, we can cross-check the above results using the well known\cite{FS1} Cauchy transform formula linking the ACP and the AICP. 
In particular, (\ref{integralrepres3}) coincides with the integral representation of the Hermite polynomial\cite{ASKEY}:
\begin{align}\label{fs0}
	\pi_{k}(s,\tau) & = (-i)^{k} \sqrt{\frac{N}{2 \pi \tau}} \int_{-\infty}^\infty q^{k} \exp \left ( -\frac{N}{2\tau} (q-is)^2 \right )\dd q .
\end{align}
and the aforementioned Cauchy transform formula reads:
\begin{align}
	\theta_N(z,\tau) & = \frac{1}{c_{N-1}^2} \int \frac{\dd s}{z-s} \pi_{N-1}(s,\tau) \exp \left (-\frac{N s^2}{2\tau} \right ), 
	\label{fs1}
\end{align}
where the constant $c_k^2 = \sqrt{\frac{2\pi \tau}{N}} \left ( \frac{\tau}{N} \right )^k k!$ is the normalization of the monic polynomials. Note that this is valid only for this particular, simplest initial condition. Analogical prescriptions for other cases are significantly more complicated{\color{MidnightBlue} \cite{DF1}}. 

We start this calculation by plugging (\ref{fs0}) into  (\ref{fs1}). After transforming the $q^{N-1}$ term into a differentiation of the exponent with respect to $s$, followed by integrating by parts, we get the result:
\begin{align}
\theta_N(z,\tau) = \sqrt{\frac{N}{2\pi\tau}}  \int^{\infty+z}_{-\infty+z}  \frac{1}{u^N} \exp{\left(-N\frac{(u-z)^2}{2\tau}\right)} \dd u. \label{rair}
\end{align}
This contour in turn can be deformed to the real axis and a half circle enclosing the pole at $0$ from above (${\rm Im} z>0$) or below (${\rm Im} z<0$) -
 in complete agreement with the results given before.

\section{Large $N$ spectral dynamics}

Let $\lambda_{i}$'s be the eigenvalues of the diffusing matrix $H(\tau)$.  
The connection between the spectral density $\rho(\lambda)\equiv \langle \sum_i \delta(\lambda-\lambda_{i})\rangle $, in the limit of $N$ going to infinity, and the averaged characteristic polynomial, is established through the so-called Green's function, defined by:
\begin{align}\label{def:g}
G(z,\tau)\equiv\frac{1}{N}\langle{\rm Tr}\left[z-H(\tau)\right]^{-1} \rangle.
\end{align}
Note that when $N$ goes to infinity, the poles of this function merge, forming a cut on the complex plane.
The link is made with the well known Sokhotski-Plemelj formula $\rho(\lambda) = \frac{1}{\pi} \lim_{\epsilon\to 0_\pm} \textrm{Im} G(\lambda \mp i \epsilon)$ and the relation
\begin{align}
\label{eq:g}
	G(z,\tau) =  \lim_{N\to\infty}\frac{1}{N} \partial_z \ln \pi_N(z,\tau).
\end{align}
Note that $f_{N}(z, \tau)\equiv \frac{1}{N} \partial_z \ln \pi_N(z,\tau)$ is the famous Cole-Hopf transform. The diffusion equation derived for the ACP in the previous section corresponds to the following Burgers equation for $f_{N}(z, \tau)$
\begin{align}\label{eq:bev}
	\partial_\tau f_N (z,\tau) + f_N(z,\tau) \partial_z f_N(z,\tau) = -\frac{1}{2N} \partial^2_{z} f_N(z,\tau),
\end{align}
in which the ``spatial" variable $z$ is complex and the role of ``viscosity" is played by $-1/N$, a negative number.
In the large $N$ limit the viscosity vanishes, $f_{N}(z,\tau)\to G(z,\tau)$ and
\begin{align}\label{eq:beiv}
	\partial_\tau G(z,\tau) + G(z,\tau) \partial_z G(z,\tau) = 0.
\end{align}
%
This may be solved by determining curves, called characteristic lines, which are labeled by $\xi$, and  along which the solution is constant and equal to the initial condition, that is $G(z,\tau)=G_0(\xi)$, where $G_{0}(z)\equiv G(z,\tau=0)$.
In this particular case, these characteristic lines are defined in the $(z,\tau)$ hyperplane by 
\begin{align}\label{def:char}
z=\xi+\tau G_{0}(\xi).
\end{align}
The result is an implicit equation for $G(z,\tau)$ which is solved under the condition, stemming from (\ref{def:g}), that in the limit of $|z|\to\infty$, the Green's function has to vanish.
Curves which the characteristic lines are tangent to are called envelopes or caustics. Their position, $x_c$, is given by the condition
\begin{align}\label{def:sh}
0=\frac{{\rm d}z}{{\rm d}\xi}\biggr\rvert_{\xi=\xi_c}
=1+\tau G'_{0}(\xi_c). 
\end{align}
Along them, the mapping between $z$ and $\xi$ ceases to be one to one, which makes the characteristic method loose its validity.

\begin{figure}[ht!]
  \centering
   \includegraphics[width=1.\textwidth]{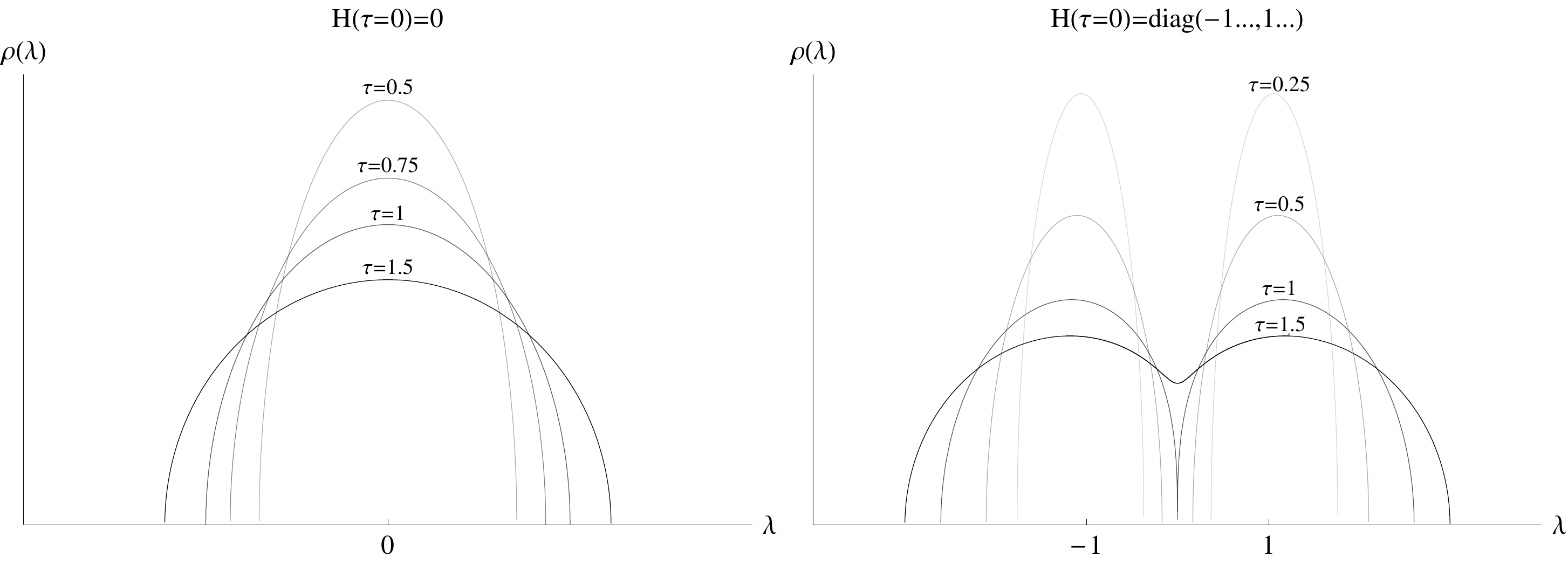}
      \caption{The above figure depicts the time evolution of the large $N$ spectral density of the evolving matrices for two scenarios that differ in the imposed initial condition. The parameter $a$ was set to one.}
     \label{fig:schemeevol}
\end{figure}

\begin{figure}[ht!]
  \centering
   \includegraphics[width=0.99\textwidth]{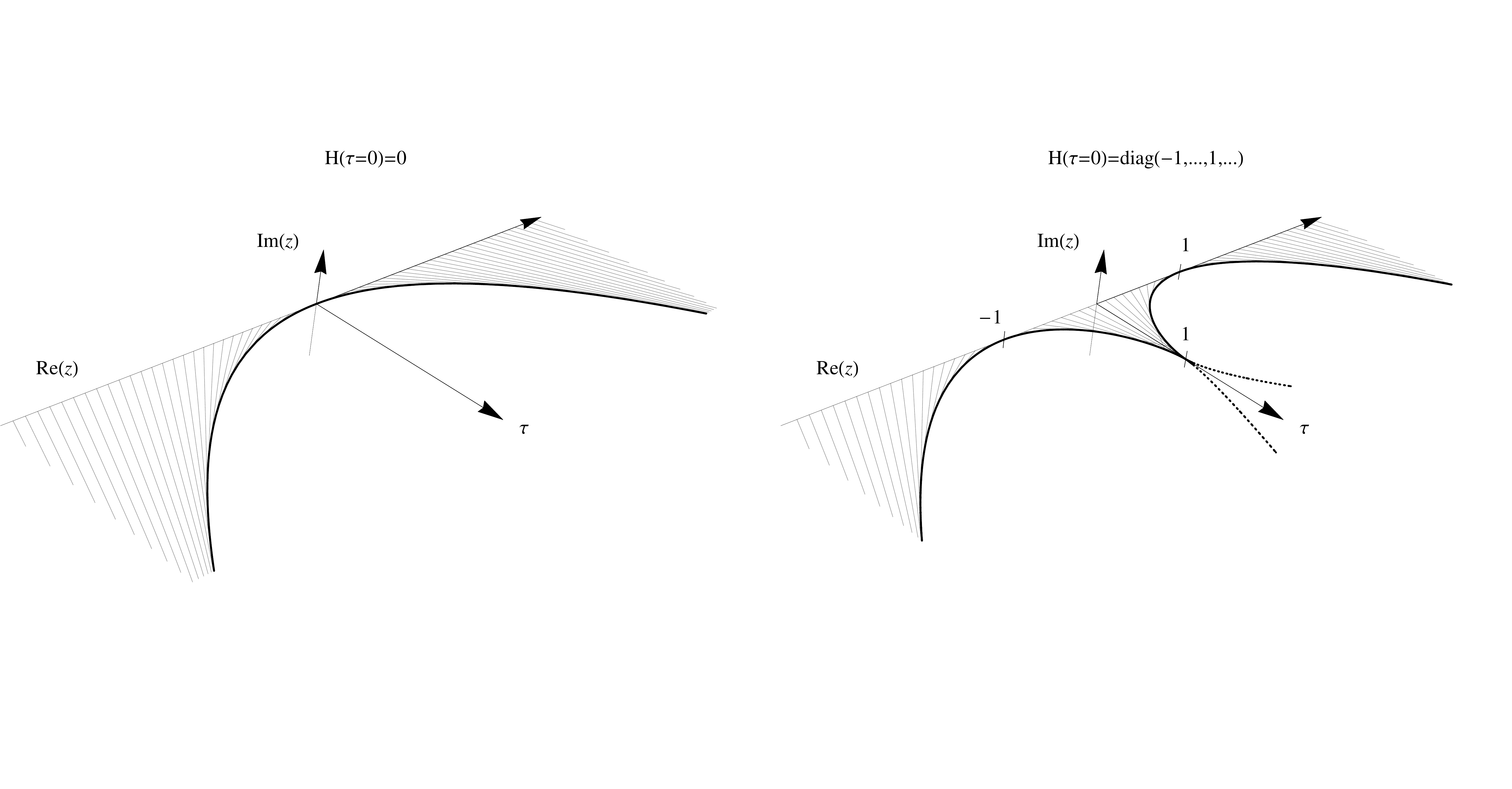}
      \caption{The thin lines are characteristics that remain real throughout their temporal evolution. They finish at the bold lines which are caustics and shocks simultaneously. The dashed bold lines are the caustics that would be formed by the strictly complex characteristics (not depicted here) if they didn't end on the branch cut.}
     \label{fig:charb}
\end{figure}

We address two types of initial conditions which are generic for this setting: $H(\tau = 0) = 0$ and $H(\tau=0) = \text{diag} (-a...,a...)$. The first one corresponds to $G_{0}=\frac{1}{z}$, while the second (for which we assume that $N$ is even) amounts to $G_{0}=\frac{1}{2(z-a)}+\frac{1}{2(z+a)}$. In the former scenario, for an infinitely large matrix, the spectrum forms a single interval throughout its evolution. In the latter, it initially occupies two separate domains of the real axis which, in turn, merge at some critical space time point.  Fig.\ref{fig:schemeevol} pictures the time evolution of the two corresponding spectral densities.
In both cases, the characteristic lines that are real at $\tau=0$, remain on the plane of ${\rm Im}z=0$ throughout the time evolution - we depict them in Fig.\ref{fig:charb}. Those that are complex, on the other hand, are symmetric under complex conjugation and never cross each other until some time, when they hit the real line and end on the cut of the Greens function. The caustics, live on the plane of ${\rm Im}( z)=0$ so long as they don't merge. Moreover, they move along the branching points of the resulting Green's function and mark, therefore, the edges of the spectra. Additionally, they constitute the positions of the shocks, curves in the $(z,\tau)$ space along which the characteristic lines have to be cut to ensure unambiguity of the solution for $G(z,\tau)$. Note finally, that if the complex characteristic lines were allowed to cross the cuts of the complex plane, they would form (in the second scenario) complex caustics evolving out of the merging point of the real ones. These are depicted by dashed lines in Fig. \ref{fig:charb}.

\section{Universal microscopic scaling} 

In this section, we inspect  the ACP and the AICP in the vicinity of the points corresponding to the edges of the spectrum of $H(\tau)$, that is near the shocks. For the size of the matrix approaching infinity, one expects that the behaviors of the ACP and the AICP do not depend on the details of the stochastic process governing the evolution - a manifestation of the so-called microscopic universality.

As some of us have demonstrated\cite{BN2}, the asymptotic behavior of the ACP can be recovered by analyzing Eq. (\ref{eq:bev}) through an expansion of $f_N$ around the positions of the shocks. Following this approach one could recover the Airy function describing the behavior of the ACP near the propagating edge. However, this method doesn't seem to be so effective when one examines a situation when two shocks collide. As we shall now see, it is more convenient in this case to return to the diffusion equation and realize that, irrespective of the initial condition, the integral representations of the ACP and AICP have the following generic structure 
\begin{align}\label{int1}
\int_{\Gamma}e^{N f(p,z,\tau)}{\rm d}p.
\end{align}
This is well suited for a steepest descent analysis in the large $N$ limit\cite{WONG}. Moreover, the saddle point condition $
\partial_{p}f(p,z,\tau)|_{p=p_i}=0 $ is equivalent to (\ref{def:char}). For the ACP, that is, we have
\begin{align}
f(p,z,\tau)=\frac{1}{N}{\rm ln}\left[\pi_0(-ip)\right]-\frac{1}{2\tau}(p-iz)^2
\end{align}
and we identify $\xi$ with $-ip_i$. For the AICP
\begin{align}
f(p,z,\tau)=\frac{1}{N}{\rm ln}\left[\theta_0(p)\right]-\frac{1}{2\tau}(p-z)^2
\end{align}
(notice that $G_{0}(p)=-\frac{1}{N}\partial_{p}{\rm ln}\left[\theta_0(p)\right]$) with $\xi$ identified as $p_i$. 
The fact that the labels of the characteristics and the saddle points are connected is clearly not coincidental. The viscid Burgers equation and the diffusion equation are equivalent trough the Cole-Hopf transform. This  induces an identification between the characteristics method used to solve the inviscid limit of the former and the saddle point method applied for the large $N$ solution of the latter. Consider approaching a caustic in $(z,\tau)$ (hyper-)plane. Through the analogy pointed out above, the merging of characteristics implies the merging of two saddle points - this will be the scenario of the first example considered below. When the caustics merge, forming a cusp, three saddle points coalesce, which will be studied subsequently.   




The final issue to resolve before engaging the calculations is the question of what precisely we mean by the ``vicinity" of the edges. If the width of the studied interval remains constant or shrinks too slowly, as the number of the eigenvalues grows to infinity, we will deal with an infinite number of eigenvalues and most of them won't ``feel" that they are ``close" to edge from the macroscopic point of view. On the other hand if the interval shrinks too fast, in the end there won't be any eigenvalues left inside the interval. This is a heuristic explanation of why the studied vicinity of the edge should have a span proportional to the average spacing of the eigenvalues near the shock. This quantity can be derived by inspecting the large $N$ limit of the spectral density, whose behavior is given, through the Sokhotski-Plemelj formula, by the Green's function. To study it around the spectral edge one expands $G$ around $\xi_c$:
\be
G_{0}(\xi)=G_{0}(\xi_c )+(\xi-\xi_c )G_{0}'(\xi_c )+\frac{1}{2}(\xi-\xi_c )^{2}G_{0}''(\xi_c )+\frac{1}{6}(\xi-\xi_c )^{3}G_{0}'''(\xi_c )+\ldots .
\ee
For any initial condition $G_0=(z-\xi)/\tau$ and thus we have $G_{0}'(\xi_c )=-{1}/{\tau}$, which gives
\be\label{zscale}
z-z_c =\frac{\tau}{k !}(\xi-\xi_c )^{k}G_{0}^{(k)}(\xi_c )+\ldots,
\ee
where $k$ in $G_{0}^{(k)}(\xi_c )$ indicates the power of the first after ($G_{0}'$), non-vanishing derivative of $G_{0}$ taken in $\xi_c$, for a given critical point. This leads to:
 \be\label{gwtausr}
G(z,\tau)\simeq G_{0}(\xi_c )+G_{0}'(\xi_c ) \left[\frac{k ! (z-z_c)}{\tau G_{0}^{(k)}(\xi_c )}\right]^{1/k}
\ee
Now, let $N_{\Delta}$ be the average number of eigenvalues located at an interval $\Delta$. We therefore have
\be  
N_{\Delta}\sim N\int^{z_c+\Delta}_{z_c}(z-z_c)^{1/k}{\rm d}z \sim N\Delta^{1+1/k}
\ee
which for fixed $N_{\Delta}$, e.g. $N_{\Delta}=1$, implies that the average eigenvalue spacing at those points is proportional to $N^{-k/(1+k)}$. 

As we shall perform the rest of the calculations using the saddle point method, let us finally show how the proper scalings arise in that framework, through the condition for the merging of the saddle points $p_i$, as was done in\cite{F1}. 
Particularly, the deviation $s$ from $z_c$ has to scale with the size of the matrix in such a way that, when $N$ grows to infinity, the value of the integrand is not concentrated at separate $p_i$'s but in the single $p_c$. This is equivalent to requiring that for such an $s$, the distance between the saddle points $p_i$ is of the order of the width of the Gaussian functions arising from expanding $f(p)$ around the respective $p_i$'s in $\exp [ N f(p) ] $. In particular the condition:
\begin{align}
\label{eq:scaling}
|p_i-p_n | \sim \left[N f''(p_j)\right]^{-1/2}, \quad (i \ne n),
\end{align}
gives the relevant order of magnitude of $s$ that is $N^{\alpha}$. We therefore set $z=z_c+s=z_c+N^\alpha \eta$ and $\eta$ is of order one. Note that $\alpha$ and $\delta$ are related through $\alpha(1+k)=-k$.
This subsequently sets the scale for the distance probed by the deviation from $p_c$ and the condition $|p_i-p_c|\sim N^{\beta}$ defines the substitution $p=p_c+N^{\beta}t$. The connection between the saddle points and characteristic lines allows us to relate $\beta$ and $\alpha$ through $k$. We see that near the critical point $|p_i-p_c|\sim |\xi-\xi_{c}|$. Using Eq. (\ref{zscale}), we thus obtain $\beta=\alpha / k=-(1+\alpha)$, which can be used as a consistency check.
In the example of subsection B the merging of the saddle points happens in a particular critical time $\tau_c$ and there exists a time scale of the order of $N^\gamma$ for which, asymptotically $p_i$ are not distinguishable. This exponent is calculated by expanding the condition for the merging of saddle points around the critical value $\tau= \tau_c + N^\gamma \kappa$.


\subsection{Soft, Airy scaling}
\begin{figure}[ht!]
  \centering
    \includegraphics[scale=.54]{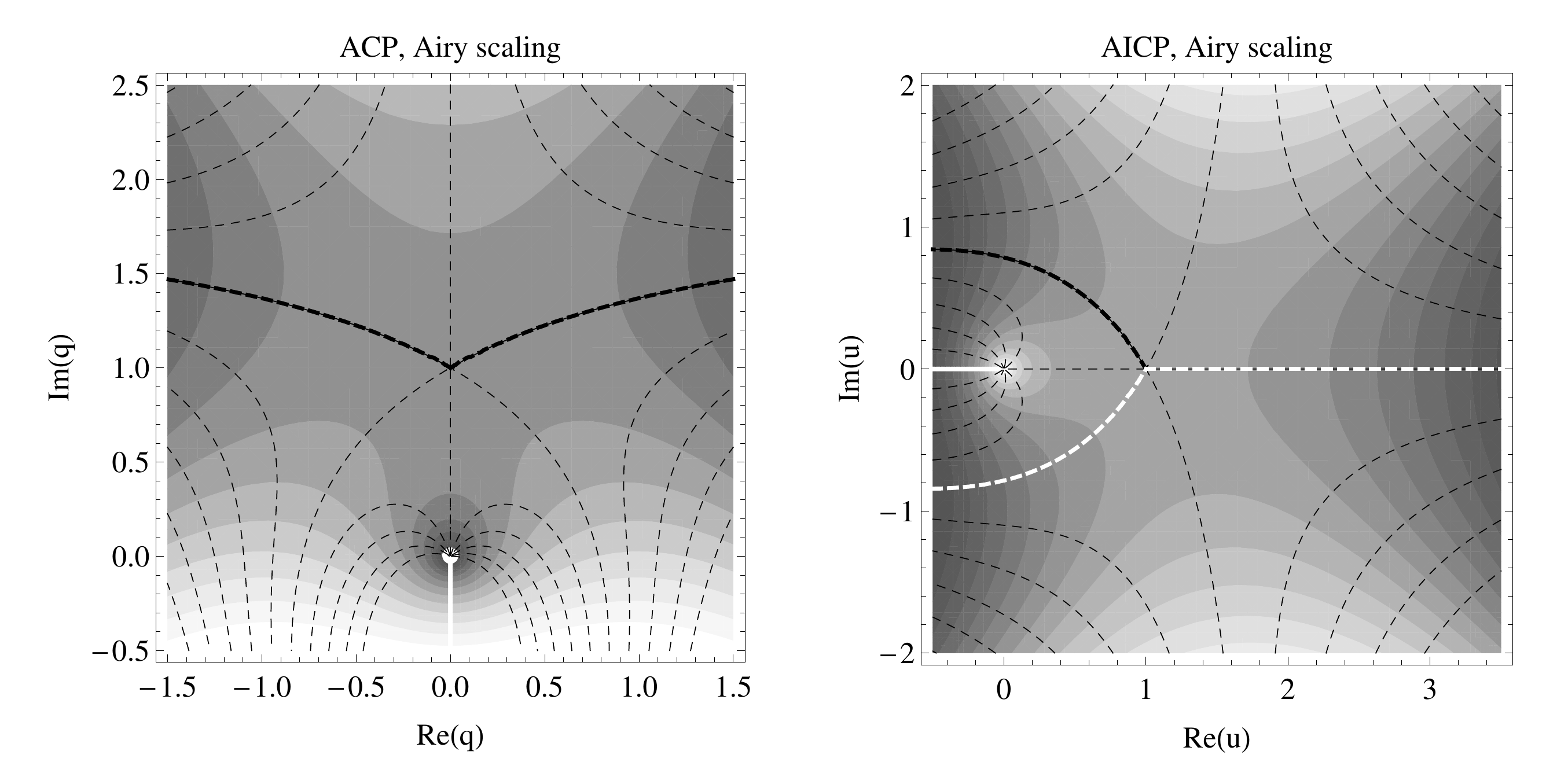}
      \caption{In the two above graphs, the gray scale gradient portrays the value of ${\rm Re}f(p)$ (growing with the brightness), whereas the dashed lines depict the curves of constant ${\rm Im}f(p)$. 
The left figure is plotted for the ACP, with $p\equiv q$, the right one for the AICP, with $p\equiv u$.  The initial condition is $H(\tau=0)=0$ and time $\tau$ is fixed to $1$ for both. 
Dashed bold curves indicate contours of integration suitable for the saddle point analysis, for the AICP, we identify the black and white line with the contours $\Gamma_+$  and $\Gamma_-$ respectively.}
     \label{fig:airysaddle1}
\end{figure}
Let us start by considering the simplest initial condition, namely $\pi_N(z,\tau=0) = z^N$, for which $H(\tau=0)$ is filled with zeros:
\begin{align}
\pi_{N}(z,\tau) & = (-i)^{N} \sqrt{\frac{N}{2 \pi \tau}} \int_{-\infty}^\infty q^{N} \exp \left ( -\frac{N}{2\tau} (q-iz)^2 \right )\dd q .
\end{align}
In this setting, the two edges of the spectrum plainly move in opposite directions along the real line (see Fig. \ref{fig:schemeevol}). We conduct a large $N$ steepest descent analysis. As a first step we set the function $f = \ln q - \frac{1}{2\tau} (q-iz)^2 $ and we compute the saddle point equation (\ref{eq:saddle}) as:
\begin{align*}
	\tau = q(q-iz).
\end{align*}
This relation has the role analogous to the hydrodynamic equation governing the characteristics as was introduced in the analysis of the Burgers equation and described above. 
The positions of the saddle points are given by
$q_\pm = \frac{1}{2} \left ( i z \pm \sqrt{4\tau - z^2} \right )$.  Their merging, at $q_c=i\sqrt{\tau}$, marks the locations of the spectral edges, in particular the right one $z_c=2\sqrt{\tau}$. Similarly, in the hydrodynamic picture, a singularity emerges when reaching the critical point $z_c$, corresponding to the crossing of the characteristic lines. 

The new contour, going trough $q_c$ is depicted (for $\tau=1$) on the left plot of figure \ref{fig:airysaddle1}. The contour deformation is constrained by two conditions: a) the real part of the function $f$ reaches a maximum along the contour and b) the imaginary part  must obey the condition $\textrm{Im} f = \textrm{const}$. By imposing the latter, we guarantee the steepest descent of $\textrm{Re} f$ upon integrating along the contour. These requirements fix uniquely the path marked in bold on the left plot of figure \ref{fig:airysaddle1}.

The scaling exponent $\alpha$ is equal to $-\frac{2}{3}$ and we have $\eta = (z-2\sqrt{\tau}) N^{2/3}$. Moreover, $\beta=-\frac{1}{3}$ and the change of variables in the integral 
is given by $t = (q-i\sqrt{\tau})N^{1/3}$. These scaling parameters were obtained from the equation (\ref{eq:scaling}) which compares the spacing between the saddle points $q_\pm$ to the width of the Gaussian approximation. The same result can be read out from the spectral density expanded around the edge of the bulk $\rho \sim \sqrt{|z-z_c|}$. Expanding the logarithm and taking the large matrix size limit yields
\begin{align}
\pi_N \left(z=2\sqrt{\tau}+ \eta N^{-2/3},\tau\right)\approx \tau^{N/2}\frac{N^{1/6}}{\sqrt{2\pi}} \exp{\left(\frac{N}{2}+\frac{\eta N^{1/3}}{\sqrt{\tau}}\right)} \textrm{Ai} \left(\frac{\eta}{\sqrt{\tau}}\right),
\end{align}
where
\begin{align}
\label{eq:airy}
	\textrm{Ai}(x) = \int_{\Gamma_0} \dd t \exp \left (\frac{it^3}{3} + itx \right ),
\end{align}
is the well-known Airy function. The contour $\Gamma_0$ is formed by the rays $-\infty \cdot e^{5i\pi/6}$ and $\infty \cdot e^{i \pi/6}$ emerging as $N$ goes to infinity. Along these rays the integral (\ref{eq:airy}) is convergent as can be seen by substituting $t \to t e^{i\pi/6}$.
%

In the case of the AICP, the initial condition used above takes the form of $\theta_N(z,\tau=0)=z^{-N}$ and (\ref{aicpsol1}) reads
\begin{align}
\theta_N(z,\tau) = \sqrt{\frac{N}{2\pi\tau}}  \int_{\Gamma_\pm} u^{-N} \exp{\left(-N\frac{(u-z)^2}{2\tau}\right)} \dd u,
\end{align}
where the contours avoid the pole at zero from above $\Gamma_+$ (for ${\rm Im}z>0$) or below $\Gamma_-$ (for ${\rm Im}z<0$), as explained in subsection IIIB. 
This time the saddle point merging for the right spectral edge occurs for $u_c=\sqrt{\tau}$. It also marks the position of the cusps of the new integration contours 
(depicted for $\tau=1$ on the second plot of figure \ref{fig:airysaddle1}).  The transformation of variables is given by $\eta = (z-2 \sqrt{\tau})N^{2/3}$ and $it = (u - \sqrt{\tau})N^{1/3}$.
Notice that the complex plane of $t$ is rotated by $\pi/2$ with respect to the one of $u$. By expanding the logarithm and taking the large matrix size limit we obtain
\begin{align} 
	\theta_N \left(z=2\sqrt{\tau}+\eta N^{-2/3},\tau\right) & \approx
i \tau^{-N/2}\frac{N^{1/6}}{\sqrt{2\pi}}
\exp{\left(-\frac{N}{2}-\frac{\eta N^{1/3}}{\sqrt{\tau}}\right)} \textrm{Ai} \left(e^{i\phi_{\pm}}\frac{\eta}{\sqrt{\tau}}\right),
\end{align}
the asymptotic behavior in terms of the Airy function, yet with its argument rotated by $\phi_+= -2\pi/3$, for ${\rm Im}z>0$, and by $\phi_-=2\pi/3$, for ${\rm Im}z<0$ in accordance 
with previous results for static matrices\cite{AF}. 
%

\subsection{Pearcey scaling}
\begin{figure}[ht!]
  \centering
    \includegraphics[scale=.54]{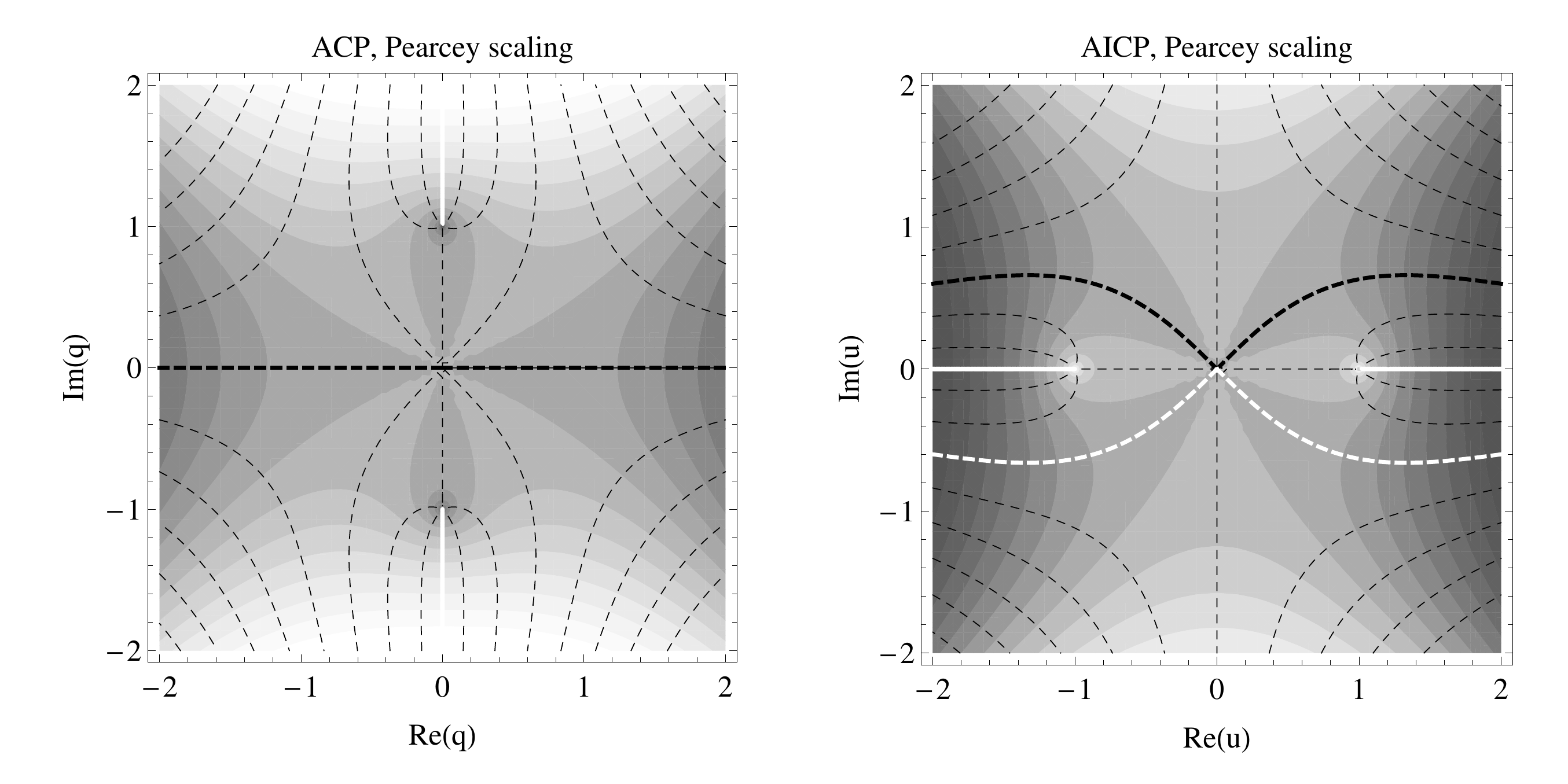}
      \caption{In the two above graphs, the gray scale gradient portrays the value of ${\rm Re}f(p)$ (growing with the brightness), whereas the dashed lines depict the curves of constant ${\rm Im}f(p)$. 
The left figure is plotted for the ACP, with $p\equiv q$, the right one for the AICP, with $p\equiv u$.  The initial condition is $H(\tau=0)={\rm diag}(1, \dots,1 , -1,\dots,-1)$ and time $\tau$ is fixed to $1$ for both. 
Dashed bold curves indicate contours of integration suitable for the saddle point analysis, for the AICP, we identify the black and white line with the contours $\Gamma_+$  and $\Gamma_-$ respectively.}
     \label{fig:pearceysaddle1}
\end{figure}
To observe a collision of the edges of the spectrum, in the large $N$ limit, one has to consider a slightly different initial condition.  Let $\pi_{N}(z,\tau=0)=(z^2 - a^2)^{N/2}$, with $N$ even. This corresponds to the initial matrix eigenvalues set to $\pm a$ with equal degeneracy $N/2$. In this case the ACP takes the form:
\begin{align}
\pi_N(z,\tau)=i^N \sqrt{\frac{N}{2\pi \tau}}\int_{-\infty}^{+\infty} \dd q \exp{\left[-\frac{N}{2\tau}(q-iz)^2 +\frac{N}{2}\log(a^2+q^2)\right]}.
\end{align}
We determine the saddle point equation:
\begin{align}
\label{eq:saddle}
	\frac{q}{q^2+a^2}- \frac{q-iz}{\tau} = 0,
\end{align}
and calculate its three solutions $q_{1,2,3}$. In this scenario the two parts of the spectra join at $z_c=0$ at $\tau_c=a^2$ and this is reflected in the saddle points merging for $q_c=0$.
The contour doesn't have to be deformed in this case as seen on the left plot of figure \ref{fig:pearceysaddle1}. 
Subsequently, the leading term of the equation (\ref{eq:scaling}) relating the distance between the solutions $q_{1,2,3}$ to the Gaussian width of the saddle point approximation is used to extract the scaling $z\sim N^{-3/4}$. Further analysis of the distance between the critical saddle point $q_c$ and the points $q_{1,2,3}$ shows that the main contribution to the integral for $z\sim N^{-3/4}$ comes from $q$ of order $N^{-1/4}$ and $\tau$ of order $N^{-1/2}$. Based on this, we set $q=t N^{-1/4}, \quad \tau = a^2 + \kappa N^{-1/2}, \quad z=\eta N^{-3/4}$.
In the limit of $N \rightarrow \infty$ we expand the logarithm arising in the exponent through the initial condition and find:
\begin{align}
\pi_N\left(z=\eta N^{-3/4},\tau \approx a^2 + \kappa N^{-1/2} \right)\approx \frac{N^{1/4}}{\sqrt{2 \pi}}(ia)^{N} \textrm P\left (\frac{\kappa}{2a^2},\frac{\eta}{a} \right ),
\end{align}
where we define the Pearcey integral by:
\begin{align}
\label{pearceydef}
\textrm P(x,y) = \int_{-\infty}^{\infty} \dd t ~ \exp \left ( -\frac{t^4}{4} + x t^2 + i t y \right ).
\end{align}
This conclusion to the analysis of the microscopic behavior of the ACP is not a surprise - the Pearcey kernel was first derived\cite{BH1} in the context of GUE matrices perturbed by a source\cite{PZJ1}, i.e. in a setting analogical to considered here, in which the sources constitute the initial condition and their critical adjustment plays the role of the critical time.

In the case of the AICP, the initial condition reads $\theta_N(z,\tau=0) = (z^2-a^2)^{-N/2}$ and the solution to the complex diffusion equation is:
\begin{align}
	\theta_N(z,\tau) = \sqrt{\frac{N}{2 \pi \tau}}\int_{\Gamma_{\pm}} \dd u ~ (u^2-a^2)^{-N/2} \exp \left ( -\frac{N}{2 \tau} (u-z)^2 \right ),
\end{align}
where $\Gamma_\pm$ denotes contours circling the poles $u_{i} = \pm a$ from above ($\Gamma_{+}$, ${\rm Im}z>0$) or from below ($\Gamma_{-}$, ${\rm Im}z<0$).
The right plot in figure \ref{fig:pearceysaddle1} depicts the types of curves these are deformed to.  
We parametrize the saddle point expansion by $\tau= a^2 + \kappa N^{-1/2}$,  $z= \eta N^{-3/4}$ and $u = e^{i\pi/4} t N^{-1/4}$.
These turn out to give:
\begin{align}
\theta_N(z=\eta N^{-3/4},\tau \approx a^2 + \kappa N^{-1/2}) \approx \frac{N^{1/4}}{\sqrt{2\pi}} (ia)^{-N}  \int_{\Gamma_{\pm}} \dd t ~ \exp \left ( -t^4/4 - \frac{i\kappa}{2a^2}t^2 + it \frac{e^{-i\pi/4} \eta}{a} \right ).	
\end{align}
We therefore obtain a Pearcey type integral along contours $\Gamma_{+}, \Gamma_{-}$ depending on the choice of sign of the imaginary part of $z$.  The former is defined by rays
with phase $\pi/2$ and $0$ whereas the later starts at $-\infty$ and after reaching zero forms a ray along a phase $-\pi/2$. 

\section{Constructing the kernel}
The ACP and the AICP are the building blocks of the matrix kernel, which in turn contains, for arbitrary $N$, all the information about the matrix model. 
Obtaining the kernel was the aim of many previous works\cite{BK1,DF1,BK2}. By making use of these results we can easily write down its form for the case of the studied diffusing matrix. We have    
\begin{align}
\label{kernel}
	K_N(x,y,\tau) = \sum_{i=0}^{N-1} \Theta_i(x,\tau) \Pi_i(y,\tau),
\end{align}
where $\Theta_i(x,\tau)$ and $\Pi_i(y,\tau)$ are defined as follows. First, let
\begin{align}
\label{teta}
	\Theta_{\vec{m}}(x,\tau) & \equiv \theta_{|\vec{m}|}^{+}(x,\tau) - \theta_{|\vec{m}|}^{-}(x,\tau) = \sqrt{\frac{N}{2\pi \tau} }\oint_{\Gamma_0} \dd u ~ \exp \left ( -N \frac{(u-x)^2}{2\tau} \right ) \theta_{0}(u) , \\\label{paj}
\Pi_{\vec{m}} (x,\tau) & \equiv \pi_{|\vec{m}|}(x,\tau) = \sqrt{\frac{N}{2\pi \tau} }\int_{-\infty}^\infty \dd q ~ \exp \left ( -N \frac{(q-ix)^2}{2\tau}  \right ) \pi_{0}(-iq).
\end{align}
Here $\theta_{0}(x) = \prod_{i=1}^d (x-a_i)^{-m_i}$, $\pi_{0}(x) = \prod_{i=1}^d (x-a_i)^{m_i}$ are the initial conditions and $\vec{m}$ is a corresponding arbitrary eigenvalue multiplicity vector. 
Moreover, the contour $\Gamma_0$ encircles all the sources $a_i$ clockwise and $\theta^+(z,\tau)$, $\theta^-(z,\tau)$ denote the different solutions of AICP diffusion equation valid for $\im z>0$ and $\im z<0$ respectively. 
Finally, the functions labeled by the index $i$ in \eqref{kernel} arise through an ordering of the multiplicities according to their increasing norm  $|\vec{m}| \equiv \sum_{j=1}^d m_j$ (see \eqref{vectors}):
\begin{align}
	\Theta_i \equiv \Theta_{\vec{n}^{(i+1)}}, \qquad \Pi_i \equiv \Pi_{\vec{n}^{(i)}}, \qquad i = 0, ..., N-1.
\end{align} 
The details concerning the construction of (\ref{kernel}) are delivered in Appendix A. It is shown in this Appendix that the Kernel for the Pearcey process that derives from the present construction is identical to that obtained by Brezin and Hikami\cite{BH1}:

\begin{align}
\label{bhclass}
	K_{BH}(x,y,\tau) = - \frac{N}{2\pi \tau}\oint_{\Gamma_0} \dd u \int_{-\infty}^\infty \dd q \frac{(-q^2-a^2)^{N/2}}{(u^2-a^2)^{N/2}} \frac{1}{u+iq} \exp \left ( -N \frac{(q-iy)^2}{2\tau} -N \frac{(u-x)^2}{2\tau}  \right ).
\end{align}


\section{Conclusions}
In this work we have studied the behavior of averaged characteristic polynomials and averaged inverse characteristic polynomials
associated with Hermitian matrices filled with entries performing Brownian motion in the space of complex numbers. A key step of our analysis was a derivation of partial differential equations governing the matrix-valued evolution. These turned out to be, independently of the initial conditions, complex heat equations with diffusion coefficients
inversely proportional to the size of the matrices, and thus provided us with integral representations of the polynomials. By using the saddle point method we were able 
to examine their so-called critical microscopical behavior, which is known to be universal. In particular, the asymptotics are driven by the Airy functions, at the edges of the spectrum, and 
by the Pearcey functions, when those edges meet. The first case holds for any frozen moment of time, however one can easily modify the free diffusion into an Ornstein-Uhlenbeck problem to obtain the Airy behaviour in the stationary limit at $\tau\to\infty$.

Finally we note, that such an analysis is also possible in the case of diffusing Wishart\cite{BNW2} or chiral\cite{CHIRBURG} matrices for which the Pearcey function is replaced by the Bessoid function 
and an additional Bessel type behavior emerges. 

\section{Acknowledgements}
MAN appreciates inspiring discussions with Nicholas Witte. PW is supported by the International PhD Projects Programme of the Foundation
for Polish Science within the European Regional Development Fund of the European Union, agreement no. MPD/2009/6 and the ETIUDA scholarship under the agreement no. UMO-2013/08/T/ST2/00105  of the National Centre of Science. MAN and JG are supported by the Grant DEC-2011/02/A/ST1/00119 of the National Centre of Science. 

\appendix
\section{The kernel structure}

To obtain \eqref{kernel}, the formula for the kernel, we first present the connection between the diffusive model considered in this paper and the matrix model with a source introduced in \cite{ZJ1}.
First, let us notice that at time $\tau$, the ensemble of the diffusing matrices $H$ is equivalent to the ensemble of matrices defined by
\begin{align}
\label{eq:h0tau}
	X_\tau = H_0 + X \sqrt{\tau},
\end{align}
where $H_0$ is the matrix initializing the stochastic process and $X$ is random and given by the complex ($\beta=2$) GUE measure
\begin{align}
	P(X) \dd X \sim \exp \left (-\frac{N}{2} \Tr X^2 \right ) \dd X.
\end{align}
Since (\ref{eq:h0tau}) is a linear transformation, we have  $X = \frac{X_\tau - H_0}{\sqrt{\tau}}$ and  $\dd X \sim \dd X_\tau$. We therefore recover a matrix model with a source in which $\tau$ is just a parameter:
\begin{align}
	P(X_\tau) \dd X_\tau \sim \exp \left (-\frac{N}{2\tau} \Tr (X_\tau - H_0)^2 \right ) \dd X_\tau.
\end{align}
$H_0$, corresponding in our formalism to the initial condition of $\tau = 0$, is the source matrix. From now on we follow closely the works on random matrices with a source\cite{BK1,DF1,BK2}. The matrix $H_0$ can be written in a diagonal form as 
$$
H_0 = \text{diag} \left ( \underbrace{a_1 ~ a_1 ~ ...}_{n_1} ; \underbrace{a_2 ~ a_2 ~ ...}_{n_2} ; ... ;\underbrace{a_d ~ ...}_{n_d}\right ),
$$
with $d$ eigenvalues $a_i$ of multiplicities $n_i$. Out of the degeneracies we form a vector $\vec{n} = (n_1,...,n_d)$ and which has a norm $|\vec{n}| \equiv \sum_{i=1}^d n_i = N$ dictated by the matrix size.
We subsequently introduce, after \cite{BK1}, the multiple orthogonal polynomials of type I and II. The functions of type I are defined on the real line through
\begin{align}
\label{tetaa}
	\Theta_{\vec{m}}(x,\tau) \equiv \theta_{|\vec{m}|}^{+}(x,\tau) - \theta_{|\vec{m}|}^{-}(x,\tau) = \sqrt{\frac{N}{2\pi \tau} }\oint_{\Gamma_0} \dd u ~ \exp \left ( -N \frac{(u-x)^2}{2\tau} \right ) \theta_{0,\vec{m}}(u) ,
\end{align}
with an arbitrary multiplicity vector $\vec{m}$, an initial condition $\theta_{0,\vec{m}}(x) = \prod_{i=1}^d (x-a_i)^{-m_i}$ and the contour $\Gamma_0$ encircling all $a_i$'s clockwise. This contour arose since the AICP was defined in \eqref{aicpsol1} by two different contours $\Gamma_-$ and $\Gamma_+$ for $\im~ z < 0$ and $\im~ z > 0$ respectively. 
%

Analogously, polynomials of type II are defined through the averaged characteristic polynomial by
\begin{align}
\label{paj2}
\Pi_{\vec{m}} (x,\tau) \equiv \pi_{|\vec{m}|}(x,\tau) = \sqrt{\frac{N}{2\pi \tau} }\int_{-\infty}^\infty \dd q ~ \exp \left ( -N \frac{(q-ix)^2}{2\tau}  \right ) \pi_{0,\vec{m}}(-iq),
\end{align}
with an initial condition $\pi_{0,\vec{m}}(x) = \prod_{i=1}^d (x-a_i)^{m_i}$. We stress the dependency of the polynomials on the multiplicity vector $\vec{m}$ of arbitrary norm $|\vec{m}| \neq N$. 
As a last step, we introduce an ordering of the vector $\vec{n}$
\begin{align*}
	\vec{n}^{(0)} & = (0,0,...,0),\\
	\vec{n}^{(1)} & = (1,0,...,0), \\
	& \vdots \\
	\vec{n}^{(n_1)} & = (n_1,0,...,0), \\
	\vec{n}^{(n_1+1)} & = (n_1,1,...,0), \\
	& \vdots \end{align*} \begin{align}\label{vectors}
		\vec{n}^{(N)} & = (n_1,n_2,...,n_d).
\end{align}
which forms a "nested" sequence increasing in norm. This sequence is exploited to compose $N$ pairs of type I and type II polynomials
\begin{align}
	\Theta_i \equiv \Theta_{\vec{n}^{(i+1)}}, \qquad \Pi_i \equiv \Pi_{\vec{n}^{(i)}}, \qquad i = 0, ..., N-1.
\end{align} 
This stack of functions forms a kernel valid for an arbitrary source $H_0$
\begin{align}
	K_N(x,y) = \sum_{i=0}^{N-1} \Theta_i(x) \Pi_i(y).
\end{align}

As an example, we consider the case of $a_1 = a, a_2 = -a$ and multiplicities $n_1 = n_2 = N/2$. We plug in the integral representations \eqref{teta} and \eqref{paj}
\begin{align}
	K_N(x,y) = \frac{N}{2\pi \tau}\oint_{\Gamma_0} \dd u \int_{-\infty}^\infty \dd q \exp \left ( -N \frac{(q-iy)^2}{2\tau} -N \frac{(u-x)^2}{2\tau}  \right ) I(q,u),
\end{align}
where the sum over the initial conditions is denoted by $I(q,u)$. In our example it is equal to
\begin{align*}
	I(q,u) & = \sum_{j=0}^{\frac{N}{2} - 1} \frac{(-iq-a)^j}{(u-a)^{j+1}} + \frac{(-iq-a)^{N/2}}{(u-a)^{N/2}}\sum_{j=0}^{\frac{N}{2} - 1} \frac{(-iq+a)^j}{(u+a)^{j+1}} = \frac{1}{u+iq} \left ( 1 - \frac{(-q^2-a^2)^{N/2}}{(u^2-a^2)^{N/2}} \right ).
\end{align*}
By noticing that, under the integral, the first term vanishes, we arrive at the formula \eqref{bhclass} given by Brezin and Hikami\cite{BH1}.


\end{document}